\documentclass[twocolumn,floatfix,preprintnumbers,amsmath,amssymb,aps,prl]{revtex4}
\usepackage{amssymb,amsmath,graphics,epsfig}
\begin{document}
\title{Laser cooling of a diatomic molecule }
\author{E.S. Shuman, J.F. Barry, and D. DeMille}
\affiliation{Department of Physics, Yale University, PO Box 208120, New Haven, CT 06520, USA}
\date{\today}

\begin{abstract}

It has been roughly three decades since laser cooling techniques produced ultracold atoms \cite{Chu98,Tannoudji98,Phillips98}, leading to rapid advances in a vast array of fields.  Unfortunately laser cooling has not yet been extended to molecules because of their complex internal structure.  However, this complexity makes molecules potentially useful for a wide range of applications \cite{Carr09}.  For example, heteronuclear molecules possess permanent electric dipole moments which lead to long-range, tunable, anisotropic dipole-dipole interactions. The combination of the dipole-dipole interaction and the precise control over molecular degrees of freedom possible at ultracold temperatures make ultracold molecules attractive candidates for use in quantum simulation of condensed matter systems \cite{Pupillo_book09} and quantum computation \cite{DeMille02}.  Also ultracold molecules could provide unique opportunities for studying chemical dynamics \cite{Balakrishnan01,Krems08} and for tests of fundamental symmetries \cite{Tarbutt09,Flambaum07,DeMille08b}.  Here we experimentally demonstrate laser cooling of the polar molecule strontium monofluoride (SrF).  Using an optical cycling scheme requiring only three lasers \cite{Shuman09},  we have observed both Sisyphus and Doppler cooling forces which have substantially reduced the transverse temperature of a SrF molecular beam.  Currently the only technique for producing ultracold molecules is by binding together ultracold alkali atoms through Feshbach resonance \cite{Ni08} or photoassociation \cite{Sage05}.  By contrast, different proposed applications for ultracold molecules require a variety of molecular energy-level structures (e.g. unpaired electronic spin \cite{Tarbutt09,Andre06,Pupillo_book09,DeMille08b}, Omega doublets\cite{Vutha10}, etc).  Our method provides a new route to ultracold temperatures for molecules.  In particular it bridges the gap between ultracold temperatures and the $\sim1$ K temperatures attainable with directly cooled molecules (e.g. with cryogenic buffer gas cooling \cite{Weinstein98} or decelerated supersonic beams \cite{Bethlem00}).  Ultimately our technique should enable the production of large samples of molecules at ultracold temperatures for species that are chemically distinct from bialkalis.

\end{abstract}

 \maketitle
Laser cooling of atoms has enabled unprecedented access to ultracold temperatures.  The power of laser cooling generally derives from the ability of certain atoms to continuously scatter photons from a laser.  For instance, Doppler laser cooling relies on small but repetitive momentum kicks resulting from the absorption of red-detuned photons counterpropagating to the motion of an atom.  Doppler cooling of an atom of mass $m>20$ amu with visible light from room temperature to ultracold temperatures requires $>10^4$ photon scatters. To scatter this many photons, an atom must have a closed cycling transition in which each photon absorption is always followed by spontaneous decay back to the initial state.

Unfortunately there are no completely closed transitions in any real physical systems, and inevitably spontaneous decays to other states occur, usually before ultracold temperatures can be reached.  Each additional populated level requires a ``repump'' laser to return population back into the main cycle, so that photon scattering can continue.   Cycling transitions requiring one or two ``repump'' lasers are common in atomic systems, but they are quite difficult to find in molecules because of their vibrational and rotational degrees of freedom.  Control over vibrational states is particularly problematic because there is no strict selection rule governing the branching ratios for decay of an excited electronic state into different vibrational levels.  Instead these branching ratios are governed by the molecule's Franck-Condon factors (FCFs), which describe the overlap of the vibrational wavefunctions for different electronic states.  For a typical molecule, the probability to return to the original vibrational level after $10^4$ photon scatters is extremely small. Furthermore, decay from a single excited state can populate up to three rotational levels per vibrational level, since rotational selection rules generally only require that $\Delta N=0,\pm 1$. (We use $N$, $v$, $F$, and $M$ as the rotational, vibrational, total angular momentum, and Zeeman quantum numbers respectively.)  Because each substantially populated level requires an individually tunable ``repump'' laser, laser cooling of a molecule can easily require so many lasers as to be experimentally challenging.

Here we experimentally demonstrate direct laser cooling of a diatomic molecule by reducing the transverse velocity spread of a cryogenic beam of strontium monofluoride (SrF) in 1D.  Cryogenic buffer gas beam sources produce highly directional beams with large fluxes for a variety of molecules, as described elsewhere \cite{Maxwell05,Patterson07}.  Our scheme for eliminating the rotational and vibrational branching, necessary to ensure optical cycling, has been described previously \cite{Shuman09}.  We briefly recount the main points in the context of the experiment reported here.  We use the $X ^2\Sigma^+\!\rightarrow\!A^2\Pi_{1/2}$ electronic transition of SrF for cycling.  Use of the first excited state $A^2\Pi_{1/2}$ ensures that no other electronic states can be populated by spontaneous decay \cite{Allouche93}.  The $A^2\Pi_{1/2}$ state has a large spontaneous decay rate, $\Gamma=2\pi\times7$ MHz \cite{Dagdigian1974}, which enables the application of strong optical forces. We have chosen SrF primarily because its favorable FCF's dictate that only three vibrational levels will be significantly populated after $10^5$ photon scatters \cite{DiRosa2004}, in principle more than sufficient for stopping molecules in our cryogenic beam.  Rotational branching is eliminated by driving an $N=1\rightarrow N^\prime=0$ type transition \cite{Stuhl08}, where the prime indicates the excited state.  Transitions of this type lead to optical pumping into dark ground-state Zeeman sublevels not excited by the laser \cite{Berkeland2002}.  For example, in this system the $F=2~ |M|=2$ sublevels are dark when driven by linear laser polarization.  We eliminate these dark states by applying a magnetic field oriented at an angle $\theta_B$ with respect to the fixed linear laser polarization, forcing the dark states to Larmor precess into bright states.  Finally, radiofrequency sidebands on the lasers address all ground state hyperfine (HFS) and spin-rotation (SR) substructure.  With this scheme only three cooling lasers (one main $\lambda_{00}=663.3$ nm pump laser and two $\lambda_{10}=686.0$ nm and $\lambda_{21}=685.4$ nm vibrational repump lasers) are required as shown in Fig. 1.  (Here $\lambda_{ij}$ is the wavelength of the $X(v=i) \rightarrow A(v'=j)$ transitions.)

\begin{figure}
\includegraphics[height=3in]
{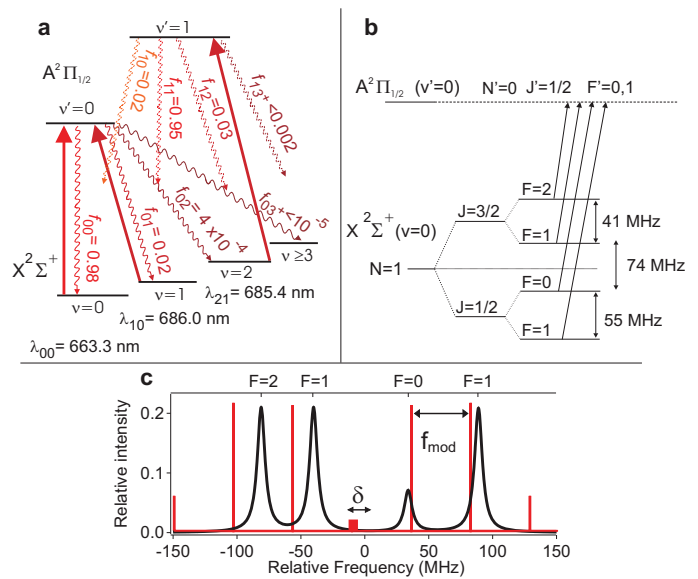} \caption{Energy level structure in SrF. a) Relevant electronic and vibrational structure.  Solid upward lines indicate laser-driven transitions in the experiment at wavelengths $\lambda_{v,v^\prime}$.  Wavy lines indicate spontaneous decays from the A($v\!=\!0$) state (solid) and the A($v\!=\!1$) state (dashed) with calculated FCFs $f_{v^\prime,v}$.  b) Relevant rotational energy levels, splittings \cite{Childs81}, and transitions (vertical arrows) in the SrF cycling scheme.  SrF has an unpaired electron spin $S=1/2$ which splits the X($N=1$) level into $J=N\pm S$ levels through SR interaction, and $^{19}$F has nuclear spin $I=1/2$ which splits $J$ into $F=J\pm1/2$ through HFS interactions.  HFS is not resolved in the A state \cite{Kandler89}. c) Laser (vertical lines) and molecular (curved) spectra for addressing all HFS ground state sublevels.  Molecular lines are shown with a power-broadened linewidth of $1.5\Gamma$ and experimentally determined line strength.  Electro-optic modulation of the laser frequency with index $m\!=\!2.6$ and frequency $f_{mod}\approx43$ MHz produces four sidebands of nearly equal strength that nearly match the HFS splitting.  The laser spectrum shown has $f_{mod}=46.4$ MHz and $\delta_p=-1.5\Gamma$, which are experimentally determined to be optimal $\lambda_{00}$ laser parameters for Doppler cooling.  The laser carrier frequency corresponding to $\delta_p=0$ is defined experimentally by the frequency (thick central line in laser spectrum) which produces maximal LIF, and $\delta_p$ corresponds to the detuning of the carrier laser frequency from this value. } \label{fig1}
\end{figure}

Use of the $N=1\rightarrow N^\prime=0$ type transition leads to a lower photon scattering rate than a traditional two-level atomic cycling transition \cite{Stuhl08,Shuman09}.  To overcome this problem we use an elongated transverse cooling region (see Supplementary Information).  The SrF beam is intersected by the three cooling laser beams at nearly right angles.  The laser beams are reflected back and forth at a slight angle so that they intersect the SrF beam $\sim 75$ times in the 15 cm long cooling region.  At the end of the cooling region the beams are nearly retroreflected, resulting in the formation of standing waves.  The combination of magnetic field remixing of Zeeman sublevels and standing waves can lead lead to Sisyphus forces in addition to Doppler forces as has been observed in atomic systems \cite{Emile93,Gupta94,Plimmer05}.  10 cm downstream from the end of the cooling region, laser induced fluorescence (LIF) is imaged to obtain the spatial distribution of the molecular beam.  This spatial distribution maps onto the velocity distribution of the molecules with a resolution of $\sim 1$ m/s, so from such images we can extract information about the velocity-dependent forces applied to the molecules, as well as the beam's transverse temperature $T$.

We find there are two cooling regimes with qualitatively different features that depend critically on the magnitude of the applied $B$ field.   In Figs. 2a and 2b we show data with $B=5$ G and 0.6 G respectively that are characteristic of these regimes.  In both regimes we find that $\theta_B$ is unimportant if $\theta_B\neq0,90^\circ$.  In all cases we observe that the total integrated LIF signal is constant to within the experimental reproducibility ($\sim5\%$), and so changes in the molecular spatial distribution accurately reflect changes in the velocity distribution of the molecular beam.  The top and bottom panels of Fig. 2a and 2b are representative molecular beam images for two different main pump laser carrier frequencies, $\delta_p=\pm 1.5 \Gamma$.  In this system there is not a single well-defined value of the detuning for the pump and repump lasers because each laser contains several sideband frequencies that each interact with multiple transitions between ground and excited states.  We define $\delta_{p(r)}=0$ experimentally by determining the pump (repump) laser carrier frequency which produces maximal LIF.  For $\delta_p=-(+) 1.5 \Gamma$ the detuning of the nearest sideband from each transition ranges from 0 to $\sim -(+)1.5 \Gamma$, indicating that the sign of $\delta_p$ corresponds to net average red (blue) detuning.  (Here and throughout the rest of the paper both vibrational repump lasers have modulation frequencies $f^{r}_{mod}=43$ MHz and $\delta_r=0$.  The main pump laser has modulation frequency $f^{p}_{mod}$ as listed in the captions.)

For a red detuned main pump laser, $\delta_p=- 1.5 \Gamma$, and $B=5$ G  we observe significant narrowing of the molecular beam  and enhancement of molecules with low transverse velocity, $v_t$, as shown in the bottom panel of Fig. 2a.  This corresponds to a reduction in the spread of $v_t$, and is a clear signature of Doppler cooling.  Also it is evident that the entire molecular beam experiences cooling forces, indicating that the cooling force is significant for all $v_t$ in the molecular beam.  The molecular beam is constrained by collimating apertures to have $v_t<v^{Beam}_{Max}\simeq 4$ m/s.  For a blue detuned main pump laser, $\delta_p=+ 1.5 \Gamma$, and $B=5$ G we observe depletion of low $v_t$ molecules and broadening of the molecular beam as shown in the top panel of Fig. 2a, as expected for Doppler heating.  Under these conditions, there is also a small but noticeable sharp feature in the center of the molecular beam, indicating some residual cooling of the remaining molecules with low $v_{t}$.

In Fig. 2b we show data characteristic of a small applied $B$ field of 0.6 G, which are strikingly different from the data in Fig. 2a.  The most significant difference between Figs. 2a and 2b is that cooling occurs for detunings of opposite sign.  For a red detuned main pump laser, $\delta_p =-1.5\Gamma$, we observe (Fig. 2b, lower panel) two relatively sharp peaks, neither of which is centered around $v_{t}=0$.  This indicates the heating of molecules with low $|v_t|$ and accumulation around two stable velocity points with $v_{t}\neq 0$.  For a blue detuned main pump laser, $\delta_p >-1.5\Gamma$, we observe (Fig. 2b, lower panel) a sharp central spike and a large enhancement of low $v_t$ molecules. This feature results from the strong cooling of molecules over a small range of $v_t$ around zero.  Meanwhile molecules with larger $|v_t|$ experience small heating forces, resulting in a very slight enhancement at large $|v_t|$.

A complete characterization of the detailed cooling forces responsible for these observations would require the solution of the optical Bloch equations for this system.  All relevant quantities (detunings, Larmor frequencies, Rabi frequencies, $\Gamma$) are the same within factors of order unity, so the full 44 level system, driven by twelve laser frequencies, each interacting with multiple levels, must be solved.  Such a calculation is beyond the scope of this work; however, the main features we have observed are common to any system with magnetically remixed dark sublevels driven by a standing wave \cite{Emile93,Gupta94,Plimmer05}.  The simplest such system is an $F=1\rightarrow0$ transition driven by a single laser frequency, and it provides substantial insight into our observations.  A linearly polarized, blue (red) detuned laser of wavelength $\lambda$ produces an AC Stark shift which attracts (repels) the $F^\prime=0,M=0$ and $F^\prime=1,M^\prime=0$ levels, while the $F=1,M=\pm1$ levels are unperturbed by the light field to first order, as shown schematically in the Supplemental Information.  In a standing wave, the $F=1,M=0$ level undergoes a spatially periodic AC Stark shift with period $\lambda/2$.  For blue (red) detuning, molecules ride up a potential hill in this level, losing (gaining) kinetic energy, before they are pumped at a rate $\gamma_{p}<\Gamma$ into the $F=1,M=\pm1$ sublevels which are dark.  If a $B$-field is applied at $\theta_B\neq0^\circ,90^\circ$ then the molecules precess from the dark $M=\pm1$ sublevels back into $M=0$ at the nodes of the standing wave at a rate $\omega_B$.  The Sisyphus force is maximized when $v_t=v^{Sis}_{Max}\sim\frac{\lambda}{4}\gamma_p$, and $\omega^{Sis}_B\sim\gamma_p$  (see Supplemental Information).  $v^{Sis}_{Max}$ defines the effective velocity range of the Sisyphus force.  Larger values of $\omega_B$ produce a much smaller Sisyphus force because the molecules are pumped back and forth between bright and dark states at random points in the standing wave.  Because the molecules spend more time in the bright states the photon scattering rate is higher, and the Doppler force is larger.  Simple arguments suggest that $\omega^{Dop}_B\sim\Gamma$ for maximum Doppler cooling forces, and that Doppler forces occur over a larger range of velocities $v_t=v^{Dop}_{Max}=\frac{\lambda \Gamma}{2\pi}$=4 m/s (see Supplemental Information).

This qualitative discussion provides substantial insight into our observations.  At low $B$ fields we expect to observe Sisyphus forces, which are characterized by cooling (heating) for blue (red) detuning for molecules with low $v_t$ as observed in Fig. 2b.  At higher $B$ fields we expect to observe Doppler forces, characterized by cooling (heating) for red (blue) detuning.  Since $v^{Beam}_{Max}\simeq v^{Dop}_{Max}$, Doppler cooling forces should affect the entire molecular beam as observed in the bottom panel of Fig. 2a.  In the Supplemental Information we provide the argument for the estimate $\gamma_p\sim\frac{\Gamma}{42}$, which yields $v^{Sis}_{Max}\sim0.2$ m/s.  This is much smaller than $v^{Beam}_{Max}\simeq v^{Dop}_{Max}\simeq4$ m/s, and is consistent with our observations.

Of course neither regime can be characterized purely by either Sisyphus or Doppler forces.  In the moderate $B$ regime residual Sisyphus forces lead to a slight additional broadening (narrowing) in the low-velocity part of the distribution for red (blue) detuning.  This residual narrowing is clear in the top panel of Fig. 2a, while the broadening is too small to observe in the bottom panel of Fig. 2a. In the low $B$ regime residual Doppler forces result in small heating (cooling) over a large range of velocities for blue (red) detuning.  This gives rise to non-zero unstable (stable) velocities $v_t = \pm v_0$, where the net force is zero and population is depleted (accumulates).  These stable points $v_t = \pm v_0$ are clear in the bottom panel of Fig. 2b.  In the top panel of Fig. 2b  the unstable velocities are clearly depleted, and there is some residual Doppler broadening in the wings of the molecular beam.  Although the arguments presented here derive from an $F=1\rightarrow F^\prime=0$ example system, they are common to any $F\rightarrow F^\prime \leq F$ system.  Furthermore the effects described here have been observed in several such atomic systems  \cite{Emile93,Gupta94,Plimmer05}.

To clearly illustrate the $B$ field dependence of the cooling force, we show the magnetic field dependence of the cooling forces for a red detuned pump laser, $\delta_p=-1.5\Gamma$, in Fig.3a.  As shown for very small B fields, the width of the molecular beam increases due to Sisyphus heating effects.  This increase is followed by a sharp decrease in the molecular beam width corresponding to Doppler cooling.  The width of the molecular beam is quite insensitive to the magnetic field over a range of intermediate magnetic field amplitudes between 2 and 6 G.  Finally at magnetic fields higher than 6 G the Doppler forces are reduced because the magnetic field artificially broadens the transitions, resulting in lower scattering rates, and lower Doppler forces.  We estimate values for $\omega_B$ in the Supplemental Information which yield $B_{Dop}\sim5$ G and $B_{Sis}\sim0.2$ G for the maximum Sisyphus and Doppler forces, in reasonable agreement with our observations.  Once again, detailed comparison would require the full solution of the optical Bloch equations for our system.

As an example of the complex features present in this system, we show the frequency dependence of the width of the SrF beam under Doppler force-dominated conditions of $B=5$ G, for various pump laser detunings $\delta_p$ in Fig.3b.  The frequency dependence of the width is substantially more complicated than that of a typical 2-level system.  As shown the force oscillates many times between heating and cooling for $-250$ MHz $<\delta_p<250$ MHz.  However, this complicated structure is amenable to a simple interpretation.  As the pump laser frequency is varied, the nearest laser frequency to each molecular transition oscillates between red-detuned (cooling) and blue-detuned (heating) with a variation that is nearly periodic in the sideband frequency.  The frequency dependence in the Sisyphus regime, shown in Fig. 3c, is somewhat more complicated, but for small detunings $-50$ MHz$<\delta_p<50$ MHz the Sisyphus force has the opposite sign as the Doppler forces as expected.  In the Supplemental Information we also show the power dependence of the Sisyphus and Doppler cooling forces which are both in reasonable agreement with expectations.

\begin{figure}
\includegraphics[height=3.0in]
{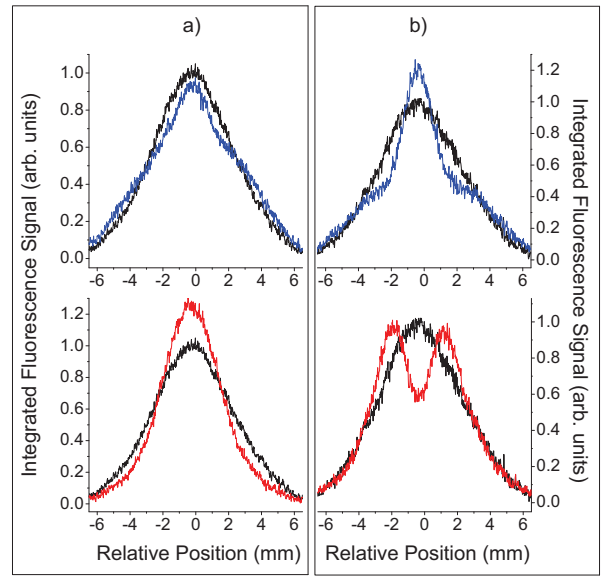} \caption{Laser cooling of SrF.  LIF in the probe region without cooling lasers in the interaction region (black curves), with cooling lasers and main pump laser red-detuned by $\delta_p=-1.5\Gamma$ (red), and with cooling lasers and main pump laser blue-detuned by $\delta_p=+1.5\Gamma$ (blue) for (a) $f^{p}_{mod}=46.4$, $B=5$ G, $\theta_B=60^\circ$, and (b) $f^{p}_{mod}=43.2$ MHz, $B=0.6$ G, $\theta_B=30^\circ$.   In (a) cooling (heating) of the beam is observed for red (blue) detuning; both are in accordance with expectations for Doppler forces.  In (b) cooling (heating) is observed for blue (red) detuning, as is expected for Sisyphus forces.  In all cases the total integrated signal is the same to within 5\%.  The blue-detuned curves systematically have the lowest total integrated signal, likely because some molecules have been been pushed outside the imaging region by the Doppler heating force.  We observe qualitatively similar behavior in both cases for pump laser modulation frequencies $42<f^{p}_{mod}<47$ MHz.  However the $f^{p}_{mod}$ values shown here were chosen because they produced the clearest Doppler and Sisyphus effects. This plot shows raw data, with no rescaling applied. } \label{fig2}
\end{figure}

\begin{figure}
\includegraphics[width=3.5in]
{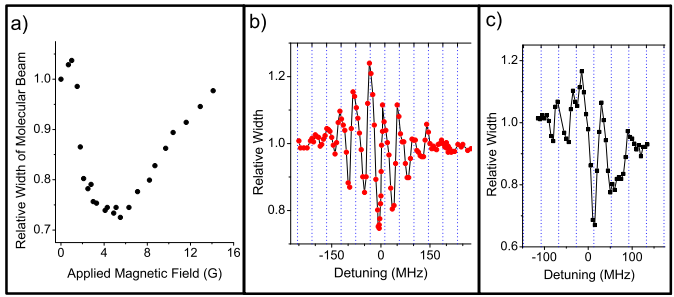} \caption{Magnetic field and frequency dependence of the cooling forces.  a) Magnetic field dependence of the cooling force. The width of the molecular beam is plotted for $f^{p}_{mod}=46.4$ MHz,  $\delta_{p}=-1.5 \Gamma$, and various $B$ fields at $\theta_B= 30^\circ$. For $B<2$ G the beam is broader due to Sisyphus forces.  At approximately $B=2$ G we see a sharp transition between Sisyphus dominated heating forces, and Doppler dominated cooling forces. For $B>6$ G the width increases as described in the main text.  Frequency dependence of the cooling forces under b) Doppler ($B=5$ G) and c) Sisyphus ($B=0.6$ G) dominated conditions. The solid lines between data points are drawn as a guide to the eye.  The vertical dashed lines are spaced by the sideband frequency of a) 46.4 MHz and b) 43.2 MHz, and illustrate the dependence of the oscillations on sideband frequency. Describing the molecular beam simply by a width is potentially problematic in the Sisyphus dominated regime where more than one velocity class are present.  Nonetheless the total width of the beam for the red (blue) detuned pump laser is larger (smaller) than the unperturbed beam width, and thus the width provides a reasonable description of the molecular beam.  The asymmetry around $\delta_p = 0$ in b) and c) underscores the complex nature of this system and is likely representative of the fact that every sideband does not have the same detuning from the nearest transition.  Therefore, the detuning of each sideband from the nearest transition is not zero for $\delta_p=0$.  Furthermore, the absolute values of the individual sideband detunings are not the same for $+\delta_p$ and $-\delta_p$.}\label{fignew}
\end{figure}

Finally, we discuss our determination of the temperature $T$ of the molecules after they are cooled. The unperturbed beam is constrained by collimating apertures to have $T_0= 50$ mK. The long interaction region prevents a precise determination of $T$ for the cooled beam, because the beam continues to expand as it experiences an imperfectly known distribution of cooling forces throughout the interaction region.  To estimate $T$, we calculate the molecular beam profile using a Monte Carlo simulation of classical particles subjected to the qualitatively expected force vs. velocity profile (see Supplemental Information).  Using these simulations we find $T\simeq T_{Sis}=300 \mu$K for the Sisyphus regime in the top panel of Fig. 2b.  We also estimate a conservative upper limit on the temperature of  $T<T_{Sis}^{Max}=5$ mK.

For the Doppler regime, we find that $T\simeq T_{Dop}=5$ mK, and  $T<T_{Dop}^{max}=15$ mK.  These values of $T$ for the Doppler regime are consistent with the final temperature expected if the molecules are subjected to $N_{sc}\approx 500-1000$ photon scatters.  This value of $N_{sc}$ agrees with expectations based on the previously observed scattering rate for this system and roughly known interaction time.  Importantly, the total integrated signal of the Doppler-cooled beam and the unperturbed beam are the same to within the level of experimental reproducibility  $(\approx 5\%)$, indicating that our cycling scheme is highly closed.

Our results have immediate implications for a number of future experiments.  For example, the 1D cooling and optical cycling demonstrated here could dramatically improve the statistical sensitivity of searches for electron electric dipole and nuclear anapole moments \cite{Tarbutt09,DeMille08b}, by providing more collimated molecular beams and enhanced detection efficiency in these experiments.  In addition, the combination of 1D cooling and a highly closed cycling transition opens the door to laser cooling of molecules in 3D.  Given the calculated FCF's, a large fraction of molecules should scatter the $\sim40,000$ photons necessary to bring a beam of SrF to a stop, and subsequently load the molecules into a trap.  Furthermore the experimentally determined loss rate of molecules in this system is $<5\%$ for $N_{sc}\sim$1000, implying that $>10\%$ can be brought to rest, given sufficient interaction time.  The laser cooling techniques presented here are limited, from a practical standpoint, to those molecules which have closed electronic transitions with diagonal FCFs and therefore require relatively few lasers.  For this reason these techniques are applicable only to a small fraction of diatomic molecules.  However because the set of diatomic molecules is very large, this subset contains a significant number of molecules\cite{DiRosa2004}.  We are aware of perhaps a dozen diatomic molecules with a wide range of internal structures that appear amenable to laser cooling with similar methods. Laser cooling such molecules to ultracold temperatures would open the door to the study of a wide variety of new physical phenomena.

\clearpage
\section{Supplementary Information}

\subsection*{1 ~~Experimental setup}

A schematic of our experiment is shown in Supplementary Fig. \ref{fig4}.  The source of SrF in our experiment is a ($2.5\!\times\!2.5\!\times\!2.5$ cm$^3$) cryogenic buffer gas cell.  The cell is attached to the 4K surface of a liquid helium-filled dewar inside a vacuum chamber held at $\sim5\times10^{-7}$ Torr.  4K helium buffer gas flows continuously into the cell through an inlet in one face of the cell.   We typically flow 5 sccm (standard cubic centimeters per minute) of helium gas into the cell which leads to a steady state density of $\sim\!10^{15}$ cm$^{-3}$.  We introduce SrF into the cell by ablating a solid SrF$_2$ target formed by pressing granular SrF$_2$ powder into a disk.  The target is ablated using 5 ns, $\sim$20 mJ pulses of 1064 nm light from a Q-switched Nd:YAG laser typically operating at 1 Hz repetition rate.  A mixture of particles including neutral SrF is ejected from the target and thermalizes to 4K after sufficient collisions with the helium buffer gas.  A 3 mm diameter exit aperture on the cell wall directly opposite the helium inlet is used to form the molecular beam.  The beam consists of SrF, helium and the other particles created by ablation.  The SrF in the beam has a measured forward velocity of $v_{||}\!=\!200\!\pm\!30$ m/s with a spread of $\Delta v_{||}\!\approx\!40$ m/s.  The beam has $\sim\!10^9$ molecules in the X($v\!=\!0$, $N\!=\!1$) state per shot.

\begin{figure}[b]
\includegraphics[width=3.5in]
{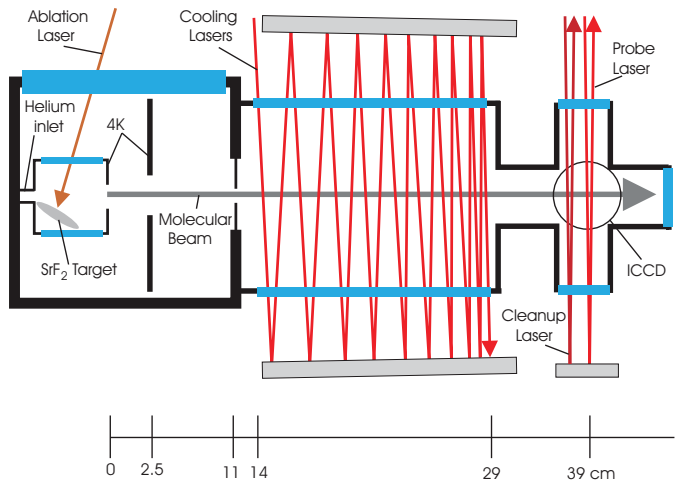} \caption{Schematic of the experimental apparatus.}\label{fig4}
\end{figure}

2.5 cm downstream from the cell, a coconut-charcoal covered copper plate with a 6.25 mm diameter hole collimates the SrF beam while also limiting the flow of helium into the rest of the apparatus.  The beam then exits the cryostat where it is further collimated by a $1.5\times1.5$ mm$^2$ aperture, resulting in an RMS transverse velocity spread $v^{RMS}_{t}=1.9$ m/s.  3 cm after the collimating aperture, the beam enters the cooling region, which is defined by 15 cm long windows.  The cooling laser beams intersect the SrF beam at nearly right angles.  The linearly polarized laser beams are reflected back and forth using 20 cm long mirrors located outside the vacuum chamber.  The windows are mounted at Brewster's angle to allow the laser beams to pass multiple times through the cooling region with minimal attenuation.  We apply $B$-fields using two pairs of rectangular coils mounted outside the cooling region.  LIF can be monitored at any position in the cooling region using a photomultiplier tube.

10 cm after the end of the cooling region, the SrF beam enters the probe region where LIF from the X$(v=0,N\!=\!1)\leftrightarrow$ A$(v^\prime=0,J^\prime\!=\!1/2)$ transition is imaged onto an intensified CCD camera.  A  ``cleanup'' $\lambda_{10}$ laser beam intersects the SrF beam between the cooling and probe regions to return any residual population in the X$(v\!=\!1)$ state back to the X($v\!=\!0$) state.  Both probe and ``cleanup'' beams are retroreflected to eliminate artificial Doppler shifts.

\subsection*{2~~Capture velocity and optimum magnetic field for Sisyphus and Doppler forces}

In this section we provide estimations of the capture velocities and optimum magnetic fields for Sisyphus and Doppler forces, which should be accurate within a factor of $\sim2$.  Once optical cycling is achieved in a system, the application of laser beams in opposite directions can lead to substantial cooling forces.  The simplest of these is the Doppler force, where the Doppler shift of a moving molecule brings it closer to or further from resonance with the laser.  Molecules closer to resonance scatter more photons, leading to a velocity-dependent force, $F_{Dop}\propto v_t$.  The Doppler force affects molecules with velocities such that the Doppler shift, $\omega_D=k v_t$, is not greater than the natural linewidth $\Gamma$, where $k=2\pi/\lambda_{00}$.  This leads to an effective velocity range for the Doppler force given by: $|v_t| \lesssim v^{Dop}_{Max}=\Gamma/k=4$ m/s.  Molecules with larger velocities than $v^{Dop}_{Max}$ experience a reduced force.

In systems with dark Zeeman sublevels, such as the SrF and $F=1\rightarrow F^\prime=0$ systems presented in the main text, optical cycling ceases as soon as the molecules are pumped into these dark states.  If a magnetic field is applied then the dark states can Larmor precess into bright states and optical cycling can continue.  The applied magnetic field must be large enough that the molecules evolve out of the dark states quickly, but it must not be so large that they precess out of the bright states before they can absorb a photon.  An estimate for the optimal $B$ field for Doppler cooling, $B_{Dop}$, can be made by equating the Larmor precession frequency, $\omega_B\sim B \mu_B g_F$, and the linewidth of the transition $\Gamma$, where $\mu_B$ is the Bohr magneton (the magnetic moment of SrF) and $g_F$ is the Land$\acute{e}$ g factor for each hyperfine level.  In SrF, $g_F\sim1$ for the hyperfine levels.  We then have $B_{Dop}\sim\frac{\Gamma}{\mu_B}=5$ G.

As described in the main text, these types of systems also give rise to Sisyphus forces if they are subjected to a standing wave.  The basic mechanism responsible for the Sisyphus effects is shown schematically in Supplementary Fig. \ref{fig3}.  Sisyphus forces are maximal for molecules that travel a distance of $\lambda/4$ in the time it takes for the molecules to be pumped into the dark states, $\frac{1}{\gamma_p}$.  In this case the laser field extracts the maximum kinetic energy from the molecules as they ride up the entire potential hill.  This leads to an effective velocity range of $v_t=v^{Sis}_{Max}\sim\frac{\lambda}{4}\gamma_p$ for Sisyphus forces.  To estimate $v^{Sis}_{Max}$ we need to determine $\gamma_p$.  In the SrF system presented in the main text, only the $F=2$, $|M|=2$ sublevels are dark in general, and a simple counting argument can give an estimate for $\gamma_p$ in this case.  If all transitions are completely saturated, then the molecules spend equal time in the 24 $v=1$ and $v=0$ ground states, and the 4 excited states.  The maximum photon scattering rate is then $\gamma^{sc}_{Max}=\frac{4}{24+4}\times\Gamma$.  According to the calculated FCF's shown in the main text, 98\% of the spontaneously emitted photons from the excited state result in decays to the $v=0$ ground state sublevels.  To first order, 1/6 of spontaneous emission events result in the population of the $|M|=2$ sublevels.  We then estimate $\gamma_p\sim\frac{\gamma^{sc}_{Max}}{6}$, or $\gamma_p\sim\frac{\Gamma}{42}$, and we find that $v^{Sis}_{Max}\sim0.2$m/s.

The previous discussion gives the optimal $v_t$ for Sisyphus forces; however, the maximum force can only be achieved for over a small range of $B$ fields.  If $B$ is too small or too large, then the molecules precess back and forth between bright and dark states at random points in the standing wave.  The net result is that after several Sisyphus cycles, the molecules have neither gained nor lost a significant amount of energy.  If on the other hand $\omega_B\simeq\gamma_p$ then molecules with $v_t\simeq v^{Sis}_{Max}$ travel $\lambda/4$ in the time it takes to precess out of the dark states.  At this point the molecules can repeat the Sisyphus process shown in Supplementary Fig. \ref{fig3}b.  Since only the $|M|=2$ sublevels are dark, we only need to consider their precession rate to estimate the optimal $B$ field for Sisyphus cooling.  For $F=2$, $g_F\simeq1/2$, and we obtain $B_{Sis}\sim\frac{\Gamma}{21\mu_B}=0.2$ G.

\subsection*{3~~Estimation of Temperature}

\begin{figure}
\includegraphics[width=3.5in]
{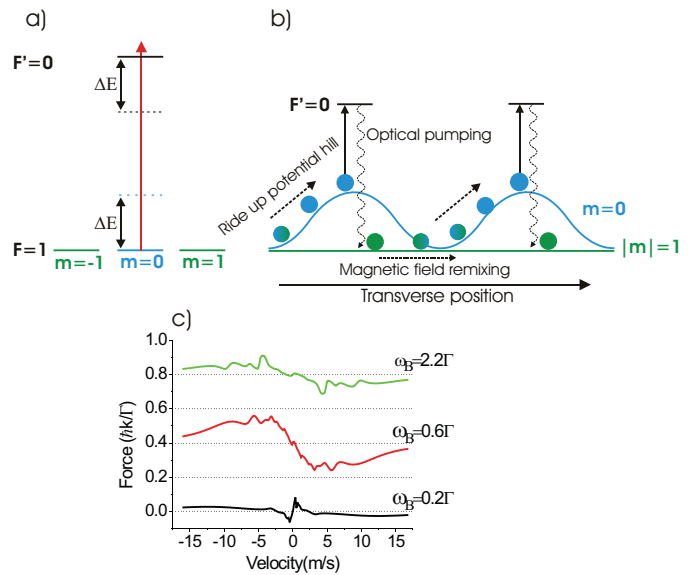} \caption{Schematic illustrating the origin of Sisyphus force in an $F=1\rightarrow F^\prime=0$ system.  (a) Zero field energy levels (solid) and AC Stark shifts (dashed) in the presence of a blue detuned, linearly polarized laser field.  (b) Energy levels and motion of molecules in a blue detuned standing wave.  As explained in the text, the molecules effectively ride continuously up potential hills, losing kinetic energy.  For red detuning the picture is reversed, with molecules gaining kinetic energy as they ride down potential hills.  For optimum Sisyphus forces, the molecules must traverse from node to antinode in the optical pumping time, $1/\gamma_p$, and the molecules must traverse from antinode to node in the magnetic field remixing time, $1/\omega_B$. (c)  Results of a semi-classical calculation of the average force on a molecule in a standing wave with Rabi frequency $\Omega_{R}=\Gamma$, and red detuning $\delta =-\Gamma$ for low (bottom), intermediate (middle), and high (top) field remixing rates $\omega_B$.  The curves are offset by 0.4 $\hbar k/\Gamma$ for clarity.  For small $\omega_B$ we observe Sisyphus heating at small velocities and small Doppler cooling at large velocities, with two stable points at $v_{t}\simeq\pm2.5$ m/s.  For higher $\omega_B$  we observe almost pure Doppler cooling forces for all velocities. For the highest $\omega_B$ the Doppler force decreases because the $B$-field broadens the transitions resulting in lower photon scattering rates. For a blue detuning of $\delta=+\Gamma$, the force is the same magnitude, but reversed in sign.  The calculation then predicts Sisyphus cooling for small $\omega_B$ and Doppler heating for higher $\omega_B$. }\label{fig3}
\end{figure}

\begin{figure}
\includegraphics[height=2.5in]
{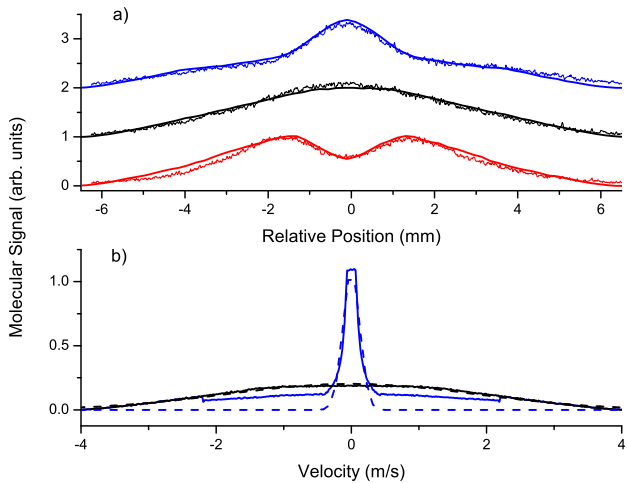} \caption{Results of a Monte Carlo simulation of the SrF beam subjected to Sisyphus type forces. a)  Experimental (noisy) and Monte Carlo simulation (clean) molecular signal for the unperturbed (black), Sisyphus cooled (blue), and Sisyphus heated (red) molecular beam. The curves are offset by one unit for clarity.  The total area of the simulated molecular beam curves are normalized to their respective experimental curve.  The difference in area between the experimental Sisyphus cooled beam and the unperturbed beam is $\sim$ 5\%. b) Velocity distribution (solid) and Gaussian fit (dashed) for the unperturbed (black) and Sisyphus-cooled (blue) molecular beam resulting from the Monte Carlo simulation of the molecular beam.  From the Gaussian fit we extract the transverse temperature $T$ of the molecular beam.}\label{fig5}
\end{figure}

As mentioned in the main text, we use a Monte Carlo simulation of classical particles subjected to the qualitatively expected Sisyphus and Doppler force versus velocity curves.  To obtain qualitative estimates of these force curves we have solved the optical Bloch equations for an $F=1\rightarrow F^\prime=0$ system in a standing wave and magnetic field.  We then compute the average force over one wavelength.  Typical results of such calculations are shown in Supplementary Fig. \ref{fig3}c, and show good agreement with the qualitative discussion in the main text.  Specifically, in the calculation we see that there are three distinct magnetic field regimes.  At low $B$ fields the calculation shows strong Sisyphus forces.  For low B fields, blue (red) detuning produces cooling (heating) forces for $|v_t|<v^{Sis}_{Max}\simeq$ 0.4 m/s.  We also see that for $|v_t|>2$ m/s there are residual Doppler forces that are of opposite sign as the Sisyphus force.  This gives rise to non-zero unstable (stable) velocities $v_t = \pm 2$ m/s, where the net force is zero and population is diminished (accumulates).  We also see that Sisyphus forces are not substantial for  $|v_t|>v^{Sis}_{Max}$, so $v^{Sis}_{Max}$ indicates the velocity extent of the Sisyphus force.

At intermediate $B$ fields the Sisyphus forces are substantially reduced, and instead the force curve is indicative of Doppler forces.  In contrast to the Sisyphus mechanism, red detuning produces Doppler cooling, while blue detuning produces Doppler heating.  Furthermore, the Doppler cooling/heating forces extend over much larger transverse velocities than the Sisyphus forces.  The Doppler forces have a broad maximum value around $v_t\sim 4$ m/s, indicating that $v^{Dop}_{Max}\sim 4$ m/s as expected.  At this intermediate $B$ field, there still remains a small residual Sisyphus force for small $v_t$ that has opposite sign to the Doppler force. At the highest $B$ fields the calculation shows reduced Doppler forces because the magnetic field artificially broadens the transitions, resulting in lower scattering rates, and lower Doppler forces.

We have used the general shape of the force curve derived from the calculation in our simulations of the molecular beam after exposure to Sisyphus and Doppler type cooling conditions.  In these simulations we have used the value of $v^{Sis}_{max}$ obtained from the $F=1\rightarrow F^\prime=0$ calculation.  In the calculation the fraction of dark states (in the absence of the remixing $B$-field) is larger than in the real SrF system; hence we take the value of $v^{Sis}_{max}$ observed in the calculations as an upper limit on the value in our experimental system.  To estimate the temperature of the Sisyphus cooled beam, we assume that the Sisyphus cooling occurs only over the last third of the cooling region (where the standing waves are most pronounced). We then adjust the magnitude of the force by an overall factor $\beta$ until the simulation matches the experimental LIF profile.

Examples of these simulations are shown in Supplementary Fig. \ref{fig5}a under Sisyphus conditions, and show good agreement with the experimental results.  The final temperature estimated in the case of Sisyphus cooling is primarily dependent on $v^{Sis}_{max}$. $v^{Sis}_{max}$ is proportional to $\gamma_{p}$, and the temperature extracted is higher for larger values of this velocity. To obtain a conservative estimate of $T$ we have used the value of $v^{Sis}_{max}$ from the $F=1\rightarrow F^\prime=0$ calculation. This value is likely larger than in the actual SrF system as shown by our estimation, so we expect that this calculation tends to overestimate $T$.  Using this simulation we find $T\simeq T_{Sis}=300 \mu$K.  However, because the full SrF system was not used to calculate the force used in the simulation, we derive a bound for the maximum value of $T$ by making two overly conservative assumptions.  First, we assume that the cooling force only occurs over a 1 cm length at the beginning of the interaction region.  Second, we allow $v^{Sis}_{cap}$ to be larger than the value found from either the $F=1\rightarrow F^\prime=0$ calculation or our estimation.  We then find the largest value of $v^{Sis}_{cap}$ which still replicates the experimental data under variation of $\beta$.  Under these very conservative assumptions we estimate $T_{Sis}^{Max}=5$ mK for $v^{Sis}_{cap}=0.8$ m/s.  This value is significantly larger than either $v^{Sis}_{cap}=0.2$ m/s from our estimation, or $v^{Sis}_{cap}=0.4$ m/s from the $F=1\rightarrow F^\prime=0$ calculation.

In the Doppler regime, we assume a uniform cooling force $F=-\alpha v$ over the whole interaction region and molecular beam velocity distribution. The overall magnitude of the force is adjusted until the simulated molecular beam width matches the width of the experimental LIF profile.  With this method we find $T\simeq T_{Dop}=5$ mK.  We also estimate a conservative upper limit by assuming this cooling force is only applied over a 1 cm length in the beginning of the cooling region; from this we find $T<T_{Dop}^{max}=15$ mK.  If we use the more realistic force vs. velocity curve shown in Supplementary Fig. \ref{fig3} c rather than $F=-\alpha v$ we obtain similar temperature estimates, and slightly better agreement between the simulation and the data.  Specifically, inclusion of the residual Sisyphus force in the Doppler force curve reproduces the small sharp feature in the center of the molecular beam in Fig. 2a of the main text.

We have also performed calculations for a system with lower states $J=3/2$ and $J=1/2$ excited to a single $J'=1/2$ upper state driven by two laser frequencies.  In this calculation we assume that each laser only drives population from the nearest $J$ level.  This calculation produced a very similar $B$-field dependence and temperature estimate as the calculation based on the simpler system.

\subsection*{4~~Power dependence of Sisyphus and Doppler cooling}

\begin{figure}
\includegraphics[height=2.5in]
{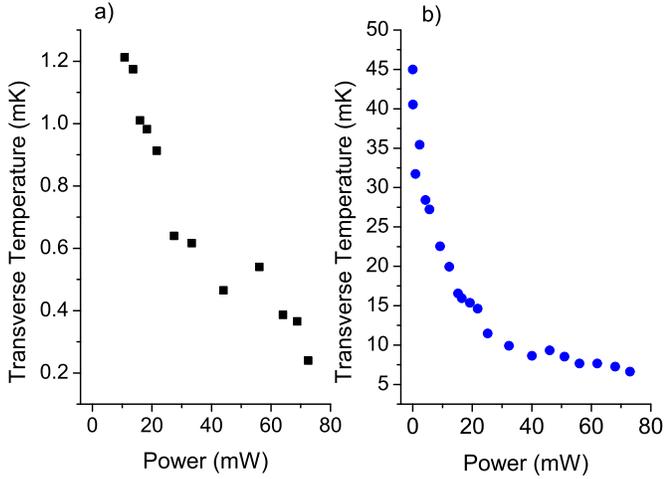} \caption{Molecular beam temperature under a) Sisyphus and b) Doppler conditions as the power in the main pump laser is varied. Doppler cooling saturates at a main laser power of approximately 30 mW, while the Sisyphus cooling continues to increase as the power is raised.  At powers below 10 mW, Sisyphus cooling is no longer observed. }\label{fig7}
\end{figure}

In this section we present additional data illustrating the dependence of the cooling forces on the laser power. In Fig. \ref{fig7} we have plotted the pump laser power dependence under both Sisyphus and Doppler cooling conditions with $\delta_p=+1.5\Gamma$ (Sisyphus cooling) and $\delta_p=-1.5\Gamma$ (Doppler cooling).  As shown in Supplementary Fig. \ref{fig7}b, the Doppler cooling force increases linearly with power until it becomes saturated.  This type of behavior is expected for Doppler cooling because at low powers the Doppler force varies as the photon scattering rate which is linear with power.  At higher powers, the photon scattering rate saturates because it is limited by the spontaneous emission rate, and larger excitation rates do not lead to larger photon scattering rates.

The Sisyphus cooling force, as shown in Fig. \ref{fig7}a, however, does not appear to saturate.  The energy extracted on each Sisyphus cycle is proportional to the AC Stark shift of the ground state.  For Rabi frequencies smaller than the detuning, the AC Stark shift varies as the intensity of the laser, while for Rabi frequencies larger than the detuning the AC Stark shift varies as the electric field of the laser. The data shown here are not sufficient to draw firm conclusions regarding a linear or quadratic power dependence; however, it is apparent that Sisyphus cooling requires more power than Doppler cooling.

\end{document}